\documentclass{article}      

\title{An Example Document}  
\author{Leslie Lamport}      
\date{January 21, 1994}      

\begin{document}
\title{Entropy and Size in HIC}

\author{   A.  Barra\~n\'on  
\footnote{ Universidad Aut\'onoma Metropolitana. Unidad Azcapotzalco.
Av. San Pablo 124, Col. Reynosa-Tamaulipas, Mexico City. email: bca@correo.azc.uam.mx } ;
 J. A.  L\'opez and J. Escamilla-Roa
\footnote{Dept. of Physics, The University of Texas at El Paso. El Paso, TX, 79968  }   } 

\def\rightmark{HIC Primitive Spinodal Decomposition}
\def\leftmark{A. Barra\~n\'on et al.}

\date{May, $5^{th}$,  2004}

\maketitle

\abstract

Distinct entropy definitions have been used to obtain an inverse
 correlation between the residual size and entropy for Heavy Ion
 Collisions. This explains the existence of several temperatures for
 different residual size bins, as reported elsewhere (Natowitz et. al., 2002). 
HIC collisions were simulated using binary interaction LATINO model
 where Pandharipande potential replicates internucleonic interaction. 
System temperature is defined as the temperature obtained when Kinetic
 Gas Theory is applied to the nucleons in the participant region. 
Fragments are detected with an Early Cluster Recognition Algorithm
 that optimizes the partitions in energy space.

\section{ Introduction}

Recent studies of the caloric curve in Heavy Ion Collisions have reported
 a system finite size effect for caloric curve limit temperatures (Natowitz y otros, 1995),
 as well as problems related to the use  double isotope thermometers for 
the sake of limit temperature estimation (Viola et. al., 1999). As a matter of fact, 
when different thermometers are used, distinct caloric curve shapes have been 
obtained such as rise-plateau-rise (Pochodzalla et. al., 1995), rise-plateau 
(Serfling et. al., 1998), or a rise-rise shape without the expected plateau 
linked to a first order phase transition (Hauger et. al., 1996). 
D$'$Agostino et. al. have developed a general protocol for the thermostatic analysis
 dealing with incomplete experimental data, showing that abnormal fluctuations
 in kinetic energy are genuine signature of a first order phase transitions in finite
 systems, delivering a limit temperature estimation close to 5.5 MeV for the 
collision of 200 particles against Au as target (D$'$Agostino et. al., 2002). 
Raduta et. al. have obtained a limit temperature in the range of 6 to 7 MeV, 
improving the computation of peaks of the microcanical caloric capacity, 
refining the primary breakup computation and including secondary particles 
emission (Raduta et. al., 2000). Borderie et. al. have applied microcanonical
 analysis of experimental data in the Fermi energy range, confirming the
 presence of a liquid-gas phase transition (Borderie et. al., 2004). 

  Beyond all these technical elements, remains the
 fundamental question of a unique caloric curve limit temperature,
 as sustained in  (Pochodzalla et. al., 1997), for differente range of
 mass, charge and energy. Nevertheless, Natowitz et. al. have 
obtained an inverse correlation between limit temperature and residual
 size (Natowitz et. al., 2002), motivating several studies on the infuence
 of the residual size and its relation with other thermodynamical 
variables such as entropy.

As a matter of fact, Gross et. al. have shown that it is posible to define
 a phase transition in finite systems on the grounds of a Statistical 
Mechanics based on Boltzmann entropy definition (Gross et. al., 2001).
 Gibbs considered the microcanonical ensamble as the fundamental one 
and the canonical as an approximation, showing that the canonical
 ensamble fails in a phase transition (J.W. Gibbs, 1902). It is not 
essential to perform a thermodynamical limit (Lebowitz, 1999) neither 
are necessary extensivity (Lieb et. al., 1998), nor concavity or 
additivity (Lavanda et. al., 1990.). Other problem with finite systems is
 the equilibrium control and the extraction of thermostatic variables 
starting from observables, in order to detect a phase transition. In the
 case of HIC, the comparation of the observed channels with the 
statistical models (Bondorf et. al., 1995) suggests that a certain 
equilibrium has been reached (Desesquelles et. al., 1998.), though 
until now no phase transition has been identified beyond any doubt
 (D$'$Agostino et. al., 2000). 

    Any Boltzmann-Gibbs equilibrium is obtained when Shannon
 entropy information is maximized in a given Fock space considering 
restrictions in the mean values of the distinct observables. Chomaz 
et. al. have developed a general definition of a phase transition based
 on annomalies in the probability distribution of the observables, 
that can be applied to non-ergodic systems, namely ensambles which
 are not like Gibbs ensambles, or even to sets of events prepared in a
 dynamical way (Chomaz et. al., 2000). Raduta et. al. have obtained 
expresions for the temperate, heat capacity and entropy second derivative
 which are  model independent  though based in the microcanonical 
multifragmentation model, using them to analyze experimental results 
and obtaining  evidence of a first order phase transition (Raduta et. al., 2001).

\section{ Methodology.}

Heavy Ion Collisions were simulated using LATINO semiclassical model
 (Barra\~n\'on et. al., 1999) where a Pandharipande potential replicates binary
 interaction. Fragments are identified with an Early Cluster Recognition 
Algorithm that optimizes configurations in energy space. Ground states
 are produced generating random configurations in phase space and gradually
 reducing the partice speed. Particles are initially confined in a parabolic
 potential and systems is gradually frozen until the theoretical binding
 energy is attained. Microscopic Persistence is used to determine the time
 at which system gets frozen. Verlet Algorithm is used to numerically 
integrate the equations of motion, using time intervals that ensure energy
 conservation in a 0.05
evidence about phase transition and critical phenomena in finite and 
transient systems,name Heavy Ion Collisions (Barra\~n\'on et. al., 2003). 
Participant region temperature is computed applying Kinetic Gas Theory
 and excitation is obtained as the energy given to the residual. System 
entropy can be computed in terms of information entropy (Ma, 1999): 
\begin{equation}
S=-\sum_{i=1}^{M} n_i ln( n_i )
\end{equation}
or with the entropy of a classical gas (Huang, 1987): 
\begin{equation}
S=log \biggl[ \bigl( \frac{1}{n} \bigr) \bigl( \frac{3T}{2} \bigr) \biggr] + S_0
\end{equation}
Fragmentation time $t_{ff}$ can be defined as the time where system 
breaks up in such a way that afterwards only some monomers are ejected.
 In orderto estimate $t_{ff}$ it is necessary to measure the simmilarities of
 partitions at different times, which can be done using the Microscopic Persistence
 Coefficient (Strachan and Dorso, 1999), which is defined as the probability
 that two particles belonging to a fragment of partition $X$ remain together
 in a fragment of partition $Y$: 

\begin{equation}
 P \bigl[ X,Y \bigr]= \frac{1}{ \sum_{fragmentos} n_i} \sum_{fragmentos} \frac{  n_i a_i } { b_i }   
\end{equation}
where $b_i$ is equal to the number of pairs of particles belonging to the cluster  $C_i$ 
of partition $X$ while $a_i $ is equal to the number of particle pairs belonging to cluster 
$C_i$ of partition $X $ that also belong to a given cluster $C'_i$ of partition $Y$. 
$n_i$ is the number of particles in cluster $C_i$.
 Therefore, fragment formation time can be defined as time where the 
Microscopic Persistence Coefficient is equal to one. At this moment, the
 biggest fragment is stable and multiplicity remains practically constant.

\section{Results}

As shown in Fig. 1, an inverse correlation is obtained between
 residual size and entropy, appying classical gas definition to the 
particles in the participant region (right). This is confirmed when
 an inverse correlation is obtained employing experimental data 
from (Natowitz et. al., 2002) (left), where limit temperature decreases
 with residual size. This inverse correlations between entropy and
 residual size was also obtained using information entropy as 
shown in Fig. 2. 

\section{Conclusions.}

As long as an inverse correlation between entropy and residual size
 is related to an experimental inverse correlation between limit
 temperature and system size, this very study encourages us to 
perform further studies on the influence of entropy on caloric 
curve limit temperature. Work supported by National Science 
Foundation (PHY-96-00038). Authors acknowledge hospitality 
from Universidad de Colima and A..B. is grateful to partial 
support from UAM-A and ready acces to the computational 
resources of the Intensive Computing Lab at UAM-Azcapotzalco.


\begin{thebibliography}{10}
\bibitem{1}
A. Barra\~n\'on et. al., 1999. Rev. Mex. Phys. {\bf 45} 110. 
\bibitem{2}
A. Barra\~n\'on et. al., 2003. Heavy Ions {\bf 17}-1 59. 
\bibitem{3}
J.P. Bondorf et. al., 1995. Phys. Rep. {\bf 257}, 133. 
\bibitem{4}
B. Borderie, 2004. Nucl.Phys. A{\bf 734} 495-503.. 
\bibitem{5}
Ph. Chomaz et. al., 2000. Preprint:arXiv:nucl-ex/0010365. 
\bibitem{6}
Ph. Chomaz et. al., 1999. Nucl. Phys. A{\bf 647}, 153. 
\bibitem{7}
M. D$'$Agostino et. al., 2002. Nucl.Phys. A{\bf 699} 795-818. 
\bibitem{8}
M. D$'$Agostino et. al., 2000. Phys.Lett. B{\bf 473} 219-225. 
\bibitem{9}
P. Desesquelles et. al., 1998. Nucl. Phys. A{\bf 633}, 547. 
\bibitem{10}
D.H.E. Gross, 2001. Preprint:arXiv:cond-mat/0105313. 
\bibitem{11}
Gibbs J.W., 1902. The Collected Works of J.Willard Gibbs, vol. II, Yale University Press, New 
Haven, 75. 
\bibitem{12}
J. A. Hauger et. al., 1996. Phys. Rev. Lett. {\bf 77}, 235. 
\bibitem{13}
Huang K., 1987. Statistical Mechanics, 2a Ed. John Wiley and Sons, New York. 
\bibitem{14}
Y.G. Ma, 1999. Phys. Rev. Lett. {\bf 83}, 3617. 
\bibitem{15}
B.H. Lavanda et. al., 1990. Foundations of Physics Letters {\bf 5}, 435. 
\bibitem{16}
J.L. Lebowitz, 1999. Rev.Mod.Phys. {\bf 71}, S346. 
\bibitem{17}
J.L. Lebowitz, 1999. Physica A{\bf 263}, 516. 
\bibitem{18}
E. H. Lieb et. al., 1998. Notices Amer. Math. Soc. {\bf 45}, 571-581 
\bibitem{19}
J. B. Natowitz et. al., 1995. Phys. Rev. C{\bf 52}, R2322. 
\bibitem{20}
J. B. Natowitz et. al., 2002. Phys. Rev. C{\bf 65}, 034618. 
\bibitem{21}
J. Pochodzalla et. al., 1995. Phys. Rev. Lett. {\bf 75}, 1040. 
\bibitem{22}
J. Pochodzalla et. al., 1997. Prog. Part. Nucl. Phys {\bf 39}, 43. 
\bibitem{23}
Al. H. Raduta et. al., 2001. Preprint:arXiv:nucl-th/0112058. 
\bibitem{24}
Al. H. Raduta et. al., 2000. Phys.Rev. C{\bf 61} 034611. 
\bibitem{25}
V. Serfling et. al., 1998. Phys. Rev. Lett. {\bf 80}, 3928. 
\bibitem{26}
A. Strachan y C. O. Dorso, 1999. Phys. Rev. C{\bf 59}, 285. 
\bibitem{27}
V. Viola et. al., 1999. Phys. Rev. C{\bf 59}, 2660.

\end{thebibliography}
\end{document}